\documentclass[a4paper, 10pt, conference]{ieeeconf}      

\IEEEoverridecommandlockouts                              

\usepackage[T1]{fontenc}
\usepackage[utf8x]{inputenc}
\usepackage{graphicx} 
\usepackage{amsmath} 
\usepackage{amssymb}  
\usepackage{cite}
\usepackage{soul, color}
\usepackage[hyphens]{url}
\usepackage{hyperref}
\usepackage{subcaption}

\title{\LARGE \bf
Deep Learning-Based Semantic Segmentation of Microscale Objects 
}

\author{Ekta U. Samani$^{1}$, Wei Guo$^{2}$, and Ashis G. Banerjee$^{3}$
\thanks{$^{1}$E. U. Samani is with the Department of Mechanical Engineering,
        University of Washington, Seattle, WA 98195, USA,
        {\tt\small ektas@uw.edu}}%
\thanks{$^{2}$W. Guo is with the Department of Industrial \& Systems Engineering, University of Washington, Seattle, WA 98195, USA,
        {\tt\small weig@uw.edu}}%
\thanks{$^{3}$A. G. Banerjee is with the Department of Industrial \& Systems Engineering and the Department of Mechanical Engineering, University of Washington, Seattle, WA 98195, USA,
        {\tt\small ashisb@uw.edu}}%
}

\begin{document}

\maketitle
\thispagestyle{empty}
\pagestyle{empty}

\begin{abstract}
Accurate estimation of the positions and shapes of microscale objects is crucial for automated imaging-guided manipulation using a non-contact technique such as optical tweezers. Perception methods that use traditional computer vision algorithms tend to fail when the manipulation environments are crowded. In this paper, we present a deep learning model for semantic segmentation of the images representing such environments. Our model successfully performs segmentation with a high mean Intersection Over Union score of 0.91.  

\end{abstract}

\section{INTRODUCTION}

Optical tweezers are widely used for non-contact manipulation of objects at the micro-scale by grasping them using tightly focused laser beams. Accurate, real-time estimation of the states (locations, sizes, etc.) of all the environment objects is necessary to automate the manipulation process. These states are typically estimated from low contrast, bright field images obtained using a charge-coupled device camera. A perception method that combines contrast enhancement, edge detection, and convolutional neural networks is proposed in \cite{rajasekaran2017accurate}. The method performs reasonably well in environments where the number of objects is limited, and successfully estimates the individual positions of the clustered objects.
However, it encounters challenges 
when a large number of objects are present in close proximity to each other. This paper provides a step toward addressing this challenge by presenting a 
deep learning-based semantic segmentation method that is capable of estimating not just the positions but also the shapes of all the objects in crowded environments. Such a capability would be useful in developing a complete situational awareness of the manipulation environments, thereby, paving the way for robust motion planning and control methods. 

\section{METHODOLOGY}
 
  We use the images from \cite{rajasekaran2017accurate} that contain multiple silica microspheres (beads) of $5\mu m$ diameter and human endothelial cells dispersed in Matrigel and Thrombin. 
  Eighty such images of resolution $640\times 480$ are used for our analysis. We define three different classes in these images, namely, the background, cells, and beads. The images are labeled using LabelMe\cite{wkentaro_2019}, a polygonal annotation tool for images. Each pixel of an image is labeled such that it belongs to one of the three defined classes. Seventy-two images are used for training and validation purposes, and eight are set aside for testing. The training and validation images are divided into 75,000 images of size $256\times 256$. These sub-images are generated by random sampling from the larger image and a randomly chosen rotation from the dihedral group. Pre-processing steps of image normalization, histogram equalization and gamma correction are performed on all the sub-images. 
  The deep learning model is then trained on 60,000 sub-images and the remaining 15,000 are used for validation.
  
  We use an encoder-decoder architecture proposed in the TGS Salt Identification challenge hosted by Kaggle\cite{kaggle}. Our model consists of Xception \cite{chollet2017xception}, pre-trained on the Imagenet database as the encoder and a ResNet-based decoder\footnote{The last decoder block does not have a concatenate layer, unlike other decoder blocks.}, as shown in Fig. 1. Intersection Over Union (IOU) is the most commonly used metric to quantify the overlap between the ground truth labels and the predicted mask. Therefore, we use the IOU score as our performance measure, which is calculated separately for each of the three classes and averaged to give a final IOU score. Lov{\'a}sz-softmax loss \cite{berman2018lovasz} is shown to be more suitable than the more generally used cross-entropy loss for optimizing the IOU metric. Hence, we choose the multi-class Lov{\'a}sz-softmax loss as our loss function. Since normalized gradient methods with constant step sizes and occasional decay perform better in deep convolutional neural networks than optimizers with adaptive step sizes such as Adam\cite{yu2017block}, we use normalized stochastic gradient descent with cosine annealing and momentum as the optimizer. We choose an initial learning rate of 0.001 and leaky rectified linear unit (ReLU) as the activation function. We do not apply any activation function to the output of our last convolutional layer. The implementation of Lov{\'a}sz-softmax loss applies a softmax activation internally to the output logits from the last layer before loss calculation.

\begin{figure*}
\centering
\includegraphics[width=1\textwidth]{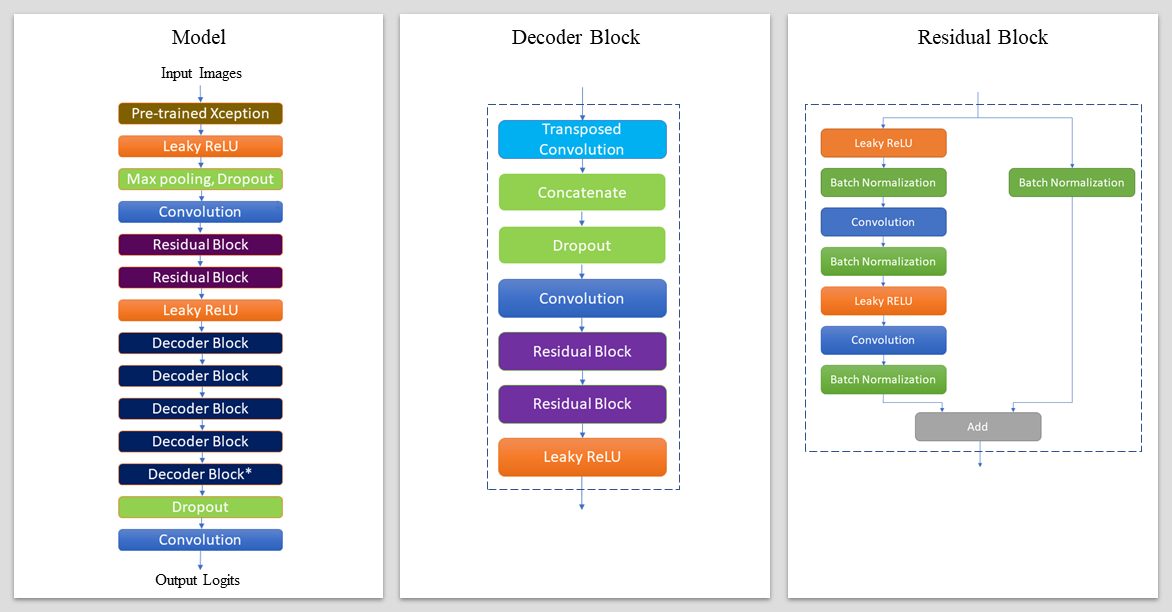}
\caption{Proposed encoder-decoder architecture with Xception as the encoder and a ResNet-based decoder. The second and third columns show the decoder and residual blocks, respectively.}
\label{figure1}
\end{figure*}

  For testing, the original image is padded with a reflection-based border padding to obtain a $768\times512 $ sized image. The padded image is divided into six non-overlapping sub-images of size $256\times 256$ that span the full image. Predictions are obtained for the six sub-images and are combined according to their respective positions to get a prediction mask for the padded image. Predictions corresponding to the border padding are discarded to obtain a final prediction mask for the original test image. We use the Opencv library to detect the contours for the cells and beads by tracing the boundaries of the segmented regions from these final prediction masks.

  We also compare the performance of our model with a fully residual convolutional neural network proposed in \cite{xie2018efficient}. The network consists of a contracting path that encodes the input to high-level features and an expanding path that decodes the features to the output mask. The contracting path consists of repeated stacks of a $3\times3$ convolution layer, a residual block and a $2\times2$ down-sampling layer. Before down-sampling, the feature map channels are doubled using a $1\times1$ convolution layer. The expanding path is similar to the contracting path, but it has up-sampling layers instead of the down-sampling layers. The down-sampling layers use mean-pooling while the up-sampling layers use bilinear interpolation. The higher resolution feature maps from contracting path are concatenated with the corresponding up-sampled feature maps in the expanding path. We use the ELU activation function and residual blocks consisting of ELU-Convolution-Dropout-ELU-Convolution-Scaling, as described in \cite{xie2018efficient}. We use the Adadelta optimizer with a learning rate of 0.0001. The network is originally designed for performing structured regression. Therefore, it uses a generalized version of the weighted square error loss function. To perform semantic segmentation using this network, we replace this loss function with the Lov{\'a}sz-softmax loss function, and we use the IOU score as the performance measure. We also change the last convolutional layer of the network to obtain three feature map channels, one for each class, in the output.
    
\section{IMPLEMENTATION AND RESULTS}

\begin{figure}
\centering
\includegraphics[width=0.5\textwidth]{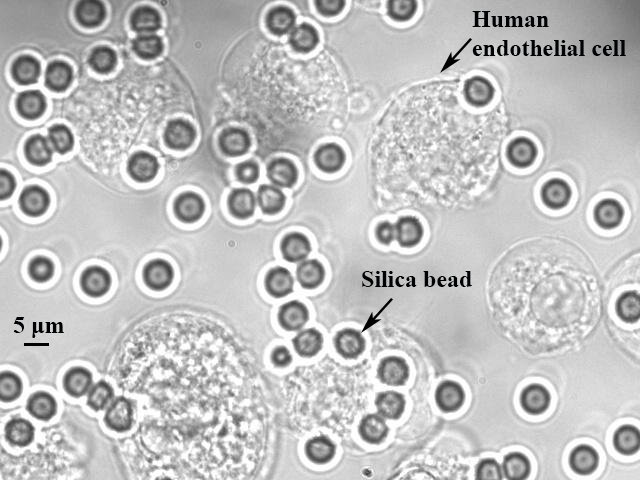}
\caption{Test image with multiple beads and cells crowded together in Matrigel medium.}
\label{figure2}
\end{figure}

\begin{figure*}[!ht]
	    \begin{minipage}[l]{1\columnwidth}
         \centering
         \includegraphics[width=1\textwidth]{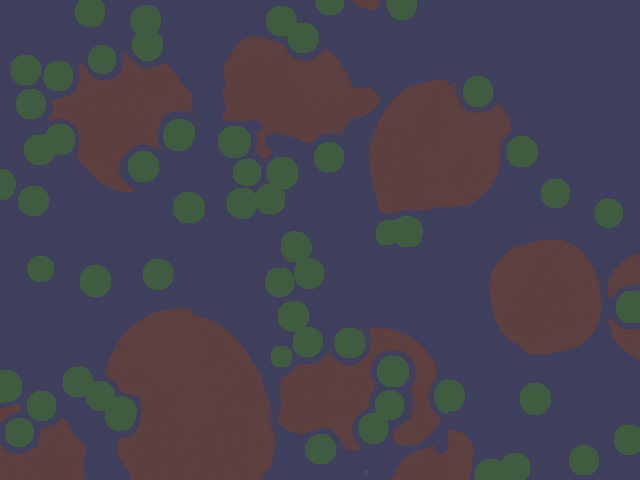}
         \subcaption{Predicted segmentation mask with IOU score of 0.89}
         \label{3a}
     \end{minipage}
     \hfill{}
     \begin{minipage}[r]{1\columnwidth}
         \centering
         \includegraphics[width=1\textwidth]{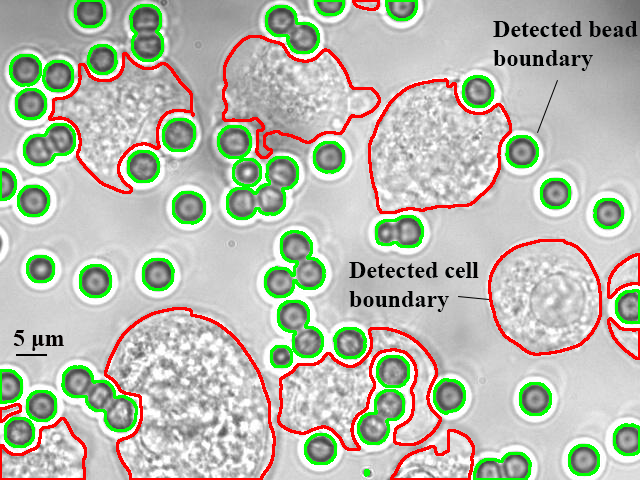}
         \subcaption{Detected boundaries overlaid on the original test image}
         \label{3b}
     \end{minipage}
     \caption{Performance of the proposed model on the test image in Fig. 2}
\end{figure*}

\begin{figure*}[!ht]
	    \begin{minipage}[l]{1\columnwidth}
         \centering
         \includegraphics[width=1\textwidth]{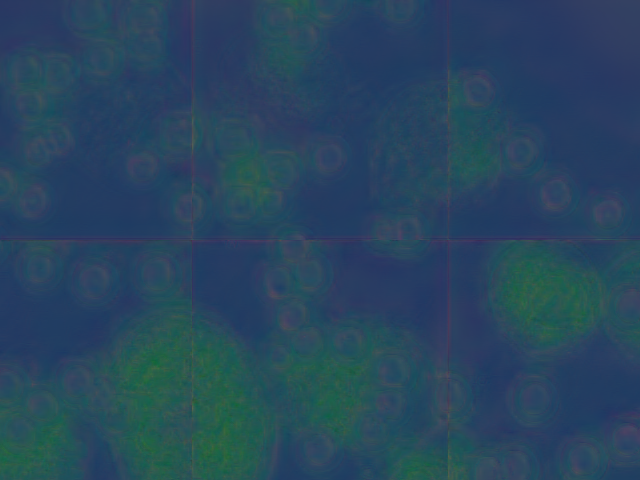}
         \subcaption{Predicted segmentation mask with IOU score of 0.27}
         \label{4a}
     \end{minipage}
     \hfill{}
     \begin{minipage}[r]{1\columnwidth}
         \centering
         \includegraphics[width=1\textwidth]{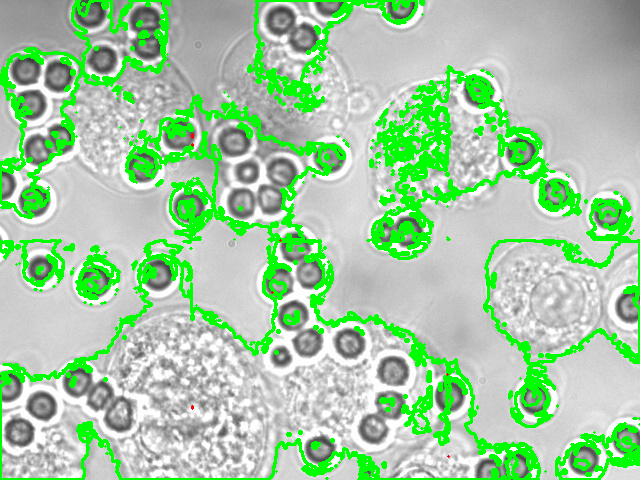}
         \subcaption{Detected boundaries overlaid on the original test image}
         \label{4b}
     \end{minipage}
     \caption{Performance of the fully residual convolutional network on the test image in Fig. 2}
\end{figure*}

All the training and testing are done on a workstation running Windows 10 operating system, equipped with a 3.7GHz 8 Core Intel Xeon W-2145 CPU, GPU ZOTAC GeForce GTX 1080 Ti, and 64 GB RAM. Our model converges with a validation IOU score of 0.97 after 16 epochs with a batch size of 8 in 48.41 hours. We obtain an IOU score of $0.91 \pm 0.02$ for the eight test images. Fig. 2 shows a typical test image with multiple cells and beads. Fig. \ref{3a} shows the corresponding predicted mask where violet color corresponds to the background, red color corresponds to the cells, and green color corresponds to the beads. Fig. \ref{3b} shows the detected contours overlaid on the original test image. Cell boundaries are red in color and bead boundaries are green in color. The model accurately segments all the cells and beads from the background. It also successfully differentiates the beads that are stuck to the cell. It is observed that most of the mislabeled pixels belong to barely visible objects on the border.

The fully residual convolutional network achieves a validation IOU score of 0.73 after 19 epochs with a batch size of 8 in 57.48 hours. We obtain an IOU score of $0.39 \pm 0.15$ for the eight test images. We observe high bias in the performance on the training set indicating that training for a longer duration may improve the predictions. Moreover, it points towards the need for a deeper network or a different architecture. We also observe high variance in the performance on the validation set despite extensive use of dropout in the network. Use of a pre-trained encoder, as in our proposed model, can help alleviate this problem to some extent. Fig. \ref{4a} shows the predicted mask and Fig. \ref{4b} shows the detected contours corresponding to the test image in Fig. 2. We observe from the segmentation mask in Fig. \ref{4a} that the model is somewhat able to identify the shape of the objects, but it is unable to differentiate between the cells and the beads. Therefore, all the detected contours in Fig. \ref{4b} are incorrectly identified as bead boundaries.

\begin{figure}
\centering
\includegraphics[width=0.5\textwidth]{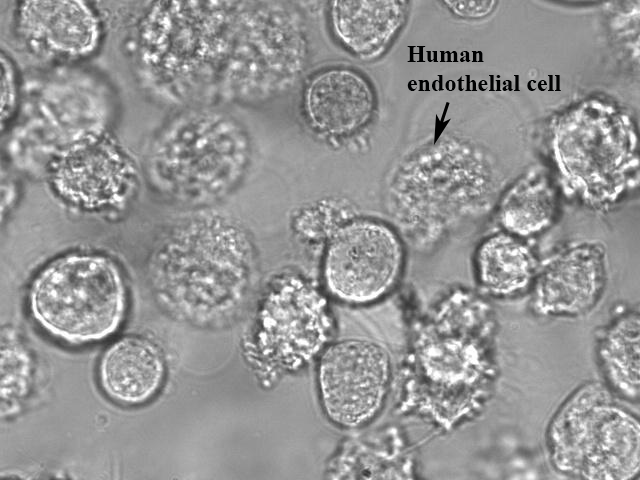}
\caption{Test image with multiple cells crowded together in poylmerizing Matrigel medium.}
\label{figure5}
\end{figure}

\begin{figure}
\centering
\includegraphics[width=0.5\textwidth]{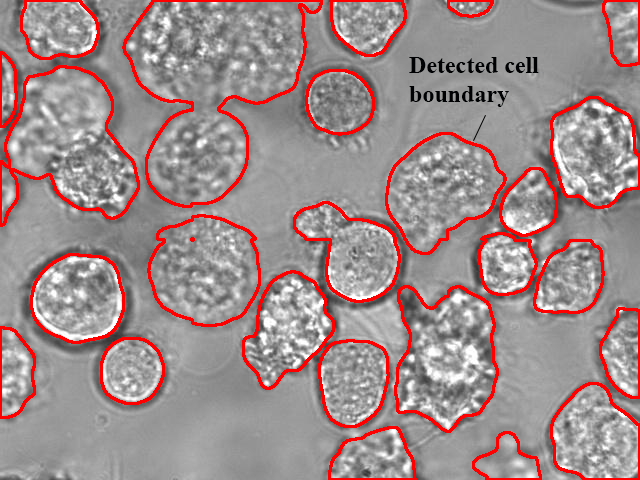}
\caption{Boundaries detected using the segmentation output (IOU score 0.92) obtained using the proposed model for the test image in Fig. 5}
\label{figure6}
\end{figure}

\begin{figure}
\centering
\includegraphics[width=0.5\textwidth]{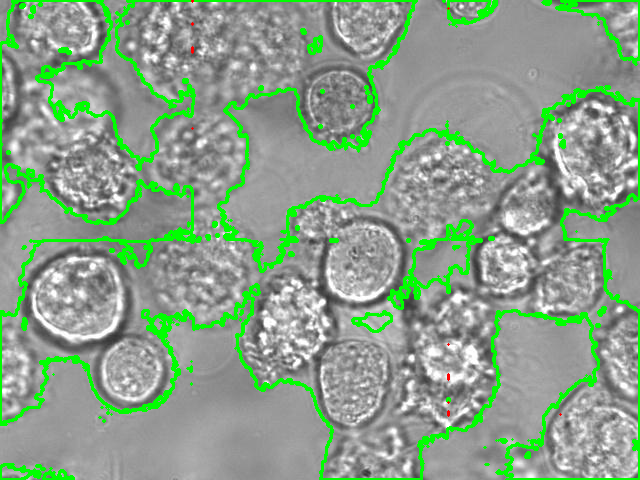}
\caption{Boundaries detected using the segmentation output (IOU score 0.27) obtained using the fully residual convolutional network for the test image in Fig. 5}
\label{figure7}
\end{figure}

Fig. \ref{figure5} shows another test image with multiple cells close together in a polymerizing Matrigel medium. Fig. \ref{figure6} shows the contours detected from the segmentation output of the proposed model. We observe that the model accurately detects all the cell boundaries in the continuously changing medium. Fig. \ref{figure7} shows the performance of the fully residual convolutional neural network on the same test image. The network fails to recognize whether the segmented regions are cells or beads. This results in incorrect identification of the detected boundaries. 

\addtolength{\textheight}{-3.4cm}

\section{CONCLUSIONS}
We present a deep learning model for accurate pixel-based semantic segmentation of micro-scale objects (cells and beads) in crowded environments. We also compare the performance of our model with a fully residual convolutional network. Our model accurately segments all the objects present in different environments. It also successfully distinguishes objects belonging to different classes that are clustered together. In the future, we plan to employ another deep learning model 
to segment the individual instances of the bead and cell classes.



\bibliographystyle{IEEEtran}
\bibliography{references}

\end{document}